\documentclass{cimento}

\usepackage{graphicx}  % got figures? uncomment this
\usepackage{amsmath}

\title{Amplitude analysis of vector-pseudoscalar final states at GlueX}
\author{A.M.~Schertz\thanks{aschertz@jlab.org} for the GlueX Collaboration}
\instlist{\item Department of Physics, Indiana University - Bloomington, Indiana, USA}
\begin{document}

\maketitle

\begin{abstract}
Excited vector mesons, some of which are predicted to include gluonic excitation in their wavefunctions, and certain classes of axial vector mesons, some of which are implicated in the decay of spin-exotic mesons, decay dominantly to vector-pseudoscalar final states. In particular, the axial vector $b_1(1235)$ decays dominantly to $\omega\pi$, and is produced copiously at the GlueX experiment. It provides a testing ground to explore strategies for amplitude analysis of further vector-pseudoscalar channels where the intermediate resonances may be harder to isolate. The formalism for amplitude analysis of $b_1^-$ production via charge exchange with a recoiling $\Delta^{++}$ is presented here, and compared with the proton recoil case. A strategy for a systematic analysis of reactions involving $\Delta^{++}$ recoil is outlined.  
\end{abstract}

\section{Introduction}

\begin{figure}[h]
    \centering
    \includegraphics[scale=0.4]{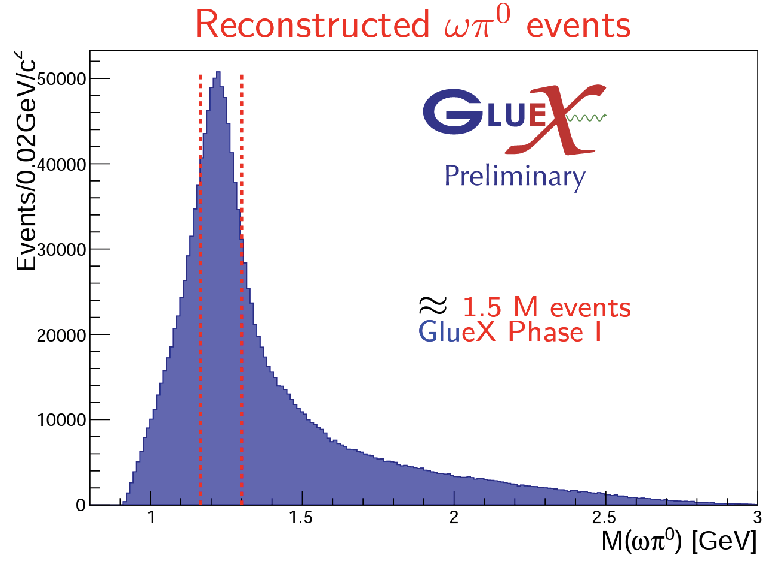}\includegraphics[scale=0.21]{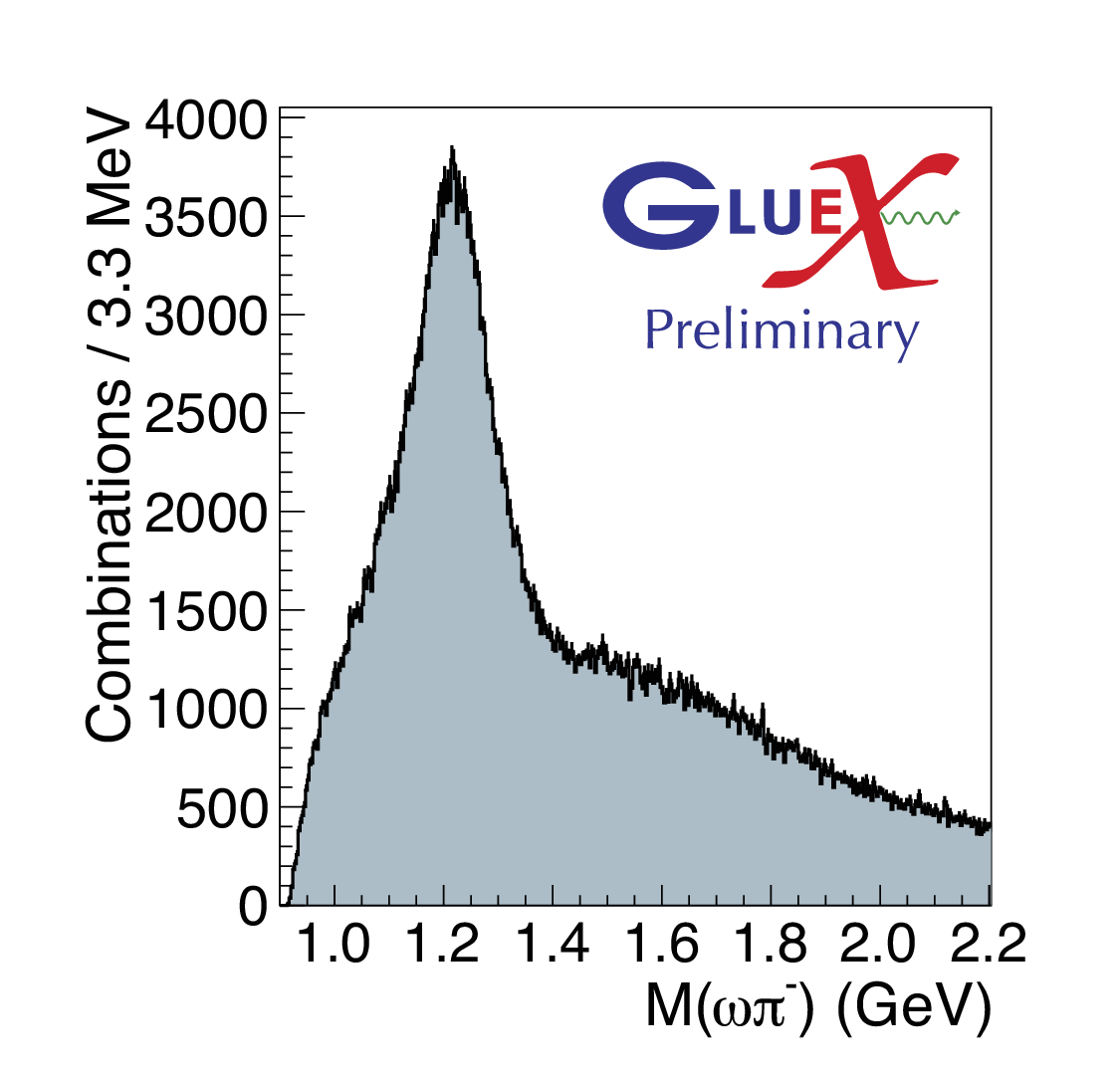}
    \caption{Left: Mass spectrum of $\omega\pi^0$ in the reaction $\gamma p \to \omega\pi^0 p$. Right: Mass spectrum of $\omega\pi^-$ in the reaction $\gamma p \to \omega\pi^+\pi^- p$. A peak at the $b_1$ mass of 1.235 GeV is clear in both plots.}
    \label{fig:massplots}
\end{figure}

The GlueX experiment~\cite{ref:alex,ref:nim} is a photoproduction experiment at Jefferson Lab in Virginia, USA with the goal of exploring the light-quark meson spectrum and searching for and understanding the nature of spin-exotic mesons, which have quantum numbers that are not allowed in the quark model. Theoretical predictions~\cite{ref:dudek} and experimental observations~\cite{ref:compass_etapi,ref:jpac_pi1,ref:bes3_etaetap} indicate that spin-exotic mesons exist, but they have yet to be observed in photoproduction. Through amplitude analyses, the quantum numbers and relative strengths of each amplitude contributing to a given final state can be extracted, allowing for identification of spin-exotic states. These proceedings focus on amplitude analysis of a resonance decaying to a vector and a pseudoscalar meson, in particular, $\omega\pi^0$ with a recoil proton, and $\omega\pi^-$ with a recoil $\Delta^{++}$ baryon. The $\omega\pi$ mass spectra for both processes are shown in Fig.~\ref{fig:massplots}. 

The most dominant contribution to the $\omega\pi$ channels is the $b_1(1235)$, whose properties are documented in Ref.~\cite{ref:pdg}. Its dominance of these channels allows for analysis of its decay properties, such as the ratio of $D$- to $S$-wave, which was measured to be $D/S = 0.269 \pm 0.009_{\tx{stat}} \pm 0.01_{\tx{sys}}$~\cite{ref:nozar}. Decay properties are not sensitive to production mechanisms, so measurement of them provides a testing ground for the amplitude analysis model used here. Vector-pseudoscalar channels also give access to excited vectors, with quantum numbers $J^{PC}=1^{--}$~\cite{ref:bes3_omegaeta,ref:snd_omegapi}, some of which are predicted to contain gluonic excitation in their wavefunctions~\cite{ref:dudek}, and to excited axial vectors, with quantum numbers $J^{PC}=1^{+-}$. 

\begin{figure}[h]
    \centering
    \includegraphics[scale=0.2]{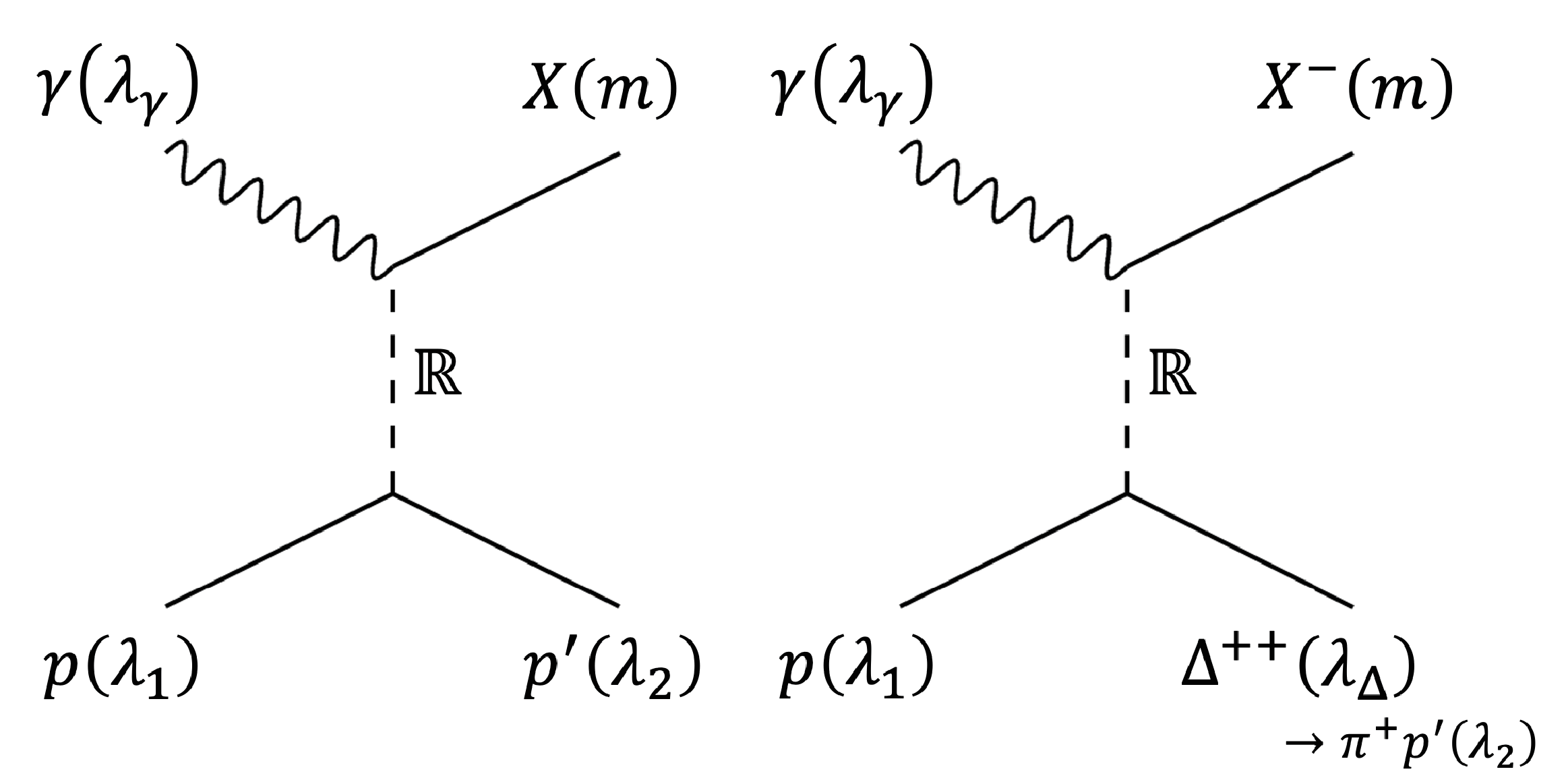}
    \caption{Photoproduction of a resonance $X$ in neutral (left) and charged (right) $t$-channel exchange. Quantities in parentheses are the helicities of the corresponding particles.}
    \label{fig:tchannel}
\end{figure}

These reactions, $\gamma p \to\omega\pi^0 p$ and $\gamma p \to\omega\pi^- \Delta^{++}$, are analyzed in the low momentum transfer region where they are assumed to proceed via exchange of a virtual particle in the \textit{t}-channel, illustrated in Fig.~\ref{fig:tchannel}. The virtual exchange particle can be electrically neutral, producing a neutral resonance recoiling off a proton, or it can carry an electric charge, producing a charged resonance recoiling off an unstable baryon, in this case a $\Delta^{++}$. Charged exchange breaks the requirement that $C$-parity be conserved, therefore certain charged mesons such as the $a^-_2(1320)$ have larger cross sections than their electrically neutral isospin partners~\cite{ref:malte}. In the case of $\omega\pi$, the neutral $\omega\pi^0$ channel shows $b_1$ production with little else, while the $\omega\pi^-$ channel still has a dominant $b_1^-$ contribution but also shows a slight shoulder around an $\omega\pi^-$ mass of 1.6 GeV. This enhancement has been previously attributed to the $\rho_3(1690)$~\cite{ref:nozar}, but confirmation of that attribution will require a rigorous amplitude analysis to properly determine its quantum numbers. The formalism for such an analysis is presented here.

\section{Formalism}
To determine the various amplitudes in a reaction such as $\gamma p\to\omega\pi^-\Delta^{++}$, an extended unbinned maximum likelihood fit is performed to the measured intensity distribution as a function of the multi-dimensional angular variables measured in experiment. 

\subsection{Proton recoil}
Appendix D of Ref.~\cite{ref:vincent_etapi} derives the intensity distribution for $\eta\pi^0$ resonances produced in the reaction $\gamma p \to \eta\pi^0 p$. By substituting an appropriate decay amplitude\footnote{here $F^j_\lambda$ is the lineshape of the resonance decaying to $\omega\pi$, and $G$ is the Dalitz function describing the $\omega$ decay},
\begin{equation} \label{eqn:zjm}
    Z^j_m(\Phi, \Omega) = e^{-i\Phi}\sum_{\lambda} F^j_\lambda D^{J_j*}_{m,\lambda} (\Omega_\omega) Y^1_\lambda (\Omega_H) G,
\end{equation}
for the phase-rotated spherical harmonics used to describe the $\eta\pi^0$ system, that intensity distribution can be rewritten in the form 
\begin{align} \nonumber
    I \propto \sum_k \bigg\{ (1-P_\gamma) & \left[ \left| \sum_{j,m} [J^P]^{(-)}_{m,k} \tx{Im} Z^j_m (\Phi,\Omega) \right|^2 + \left| \sum_{j,m} [J^P]^{(+)}_{m,k} \tx{Re} Z^j_m (\Phi,\Omega) \right|^2 \right] \\
    + (1+P_\gamma) & \left[ \left| \sum_{j,m} [J^P]^{(+)}_{m,k} \tx{Im} Z^j_m (\Phi,\Omega) \right|^2 + \left| \sum_{j,m} [J^P]^{(-)}_{m,k} \tx{Re} Z^j_m (\Phi,\Omega) \right|^2 \right] \bigg\}, \label{eqn:intenStable}
\end{align}
which is suitable for vector-pseudoscalar analysis. The quantum numbers $J^P$ of the resonance are indexed by $j$. The intensity is expressed in a basis of definite reflectivity, $\varepsilon = \pm 1$. 
In the high-energy limit for $t$-channel processes, $\varepsilon$ is equal to the naturality, $\tau = P(-1)^J$, of the virtual exchange particle.
The production amplitudes $[J^P]^{(\varepsilon)}_{m,k}$ are estimated by fitting $I(\Phi,\Omega)$ to the data. 
Notably, the recoil proton only enters into this formulation via the helicity flip between the target and recoil proton, $k$. This will be discussed further in Sec.~\ref{sec:k}. 

\subsection{Unstable $\Delta^{++}$ recoil}
The intensity function for a vector-pseudoscalar system recoiling off a $\Delta^{++}$ is
\begin{align} \nonumber
    I \propto \sum_{\lambda_2} \Bigg\{  (1-P_\gamma) & \left| \sum_{j,m,\lambda_\Delta} D^{3/2*}_{\lambda_\Delta,\lambda_2} (\Omega_p) \left( [J^P]^{(+)}_{m,\lambda_\Delta} \tx{Re} Z^j_m (\Phi,\Omega) + i [J^P]^{(-)}_{m,\lambda_\Delta} \tx{Im} Z^j_m (\Phi,\Omega) \right) \right|^2 \\ \nonumber
    + (1-P_\gamma) & \left| \sum_{j,m,\lambda_\Delta} D^{3/2}_{\lambda_\Delta,\lambda_2} (\Omega_p) \left( [J^P]^{(+)}_{m,\lambda_\Delta} \tx{Re} Z^j_m (\Phi,\Omega) - i [J^P]^{(-)}_{m,\lambda_\Delta} \tx{Im} Z^j_m (\Phi,\Omega) \right) \right|^2 \\ \nonumber
    + (1+P_\gamma) & \left| \sum_{j,m,\lambda_\Delta} D^{3/2*}_{\lambda_\Delta,\lambda_2} (\Omega_p) \left( [J^P]^{(+)}_{m,\lambda_\Delta} \tx{Im} Z^j_m (\Phi,\Omega) - i [J^P]^{(-)}_{m,\lambda_\Delta} \tx{Re} Z^j_m (\Phi,\Omega) \right) \right|^2 \\
    + (1+P_\gamma) & \left| \sum_{j,m,\lambda_\Delta}  D^{3/2}_{\lambda_\Delta,\lambda_2} (\Omega_p) \left( [J^P]^{(+)}_{m,\lambda_\Delta} \tx{Im} Z^j_m (\Phi,\Omega) + i [J^P]^{(-)}_{m,\lambda_\Delta} \tx{Re} Z^j_m (\Phi,\Omega) \right) \right|^2 \Bigg\}\label{eqn:intenUnstable}
\end{align}
where the additional Wigner D-function describes the $\Delta^{++}\to p\pi^+$ decay, and amplitudes with opposite reflectivities appear in the same coherent sum, unlike in the case with a stable recoil particle.

\section{Implications}
\subsection{Reflectivity interference}
When the recoil particle is stable, amplitudes with opposite reflectivities appear in separate coherent sums in the intensity function and cannot interfere with each other. In other words there is no interference between amplitudes produced by natural and unnatural exchange, and the notion of independent cross sections for natural and unnatural exchange exists, as are measured for the $a_2(1320)$ in the reaction $\gamma p \to \eta\pi^0 p$ in Ref.~\cite{ref:malte}. When the recoil particle is unstable, inclusion of its decay in the intensity function leads to positive and negative reflectivity amplitudes appearing in the same coherent sum. Since opposite reflectivity amplitudes add coherently, it is possible that interference between them may contribute to the intensity, and it does not make sense to extract separate cross sections for natural and unnatural parity exchange in the case of a polarized beam. 
\subsection{Separation of baryon spin flip} \label{sec:k}
In $t$-channel exchanges such as these, a quantity that is considered is the helicity flip, $k$, between the target proton and the recoil baryon. In the case of a recoil proton, $k$ can take on two possible values, related to proton helicity flip and non-flip, but in measurements taken at GlueX, it is ambiguous which value of $k$ corresponds to flip or non-flip, since it does not appear in the decay amplitude.  This quantity is summed over incoherently in the intensity function. If the recoil baryon is an unstable $\Delta^{++}$, there are four possible values of $k$, one for each value of $\Delta^{++}$ helicity, $\lambda_\Delta$, which is summed coherently in the intensity function. This coherent sum and the fact that the decay amplitudes used in the intensity function have an explicit dependence on $\lambda_\Delta$, mean that if the recoil particle is unstable, amplitudes with different values of $k$ are distinguishable and it is in principle possible to determine which, if any, is dominant.
\subsection{Analysis strategy}
In parallel with the analysis of $\gamma p\to\omega\pi^-\Delta^{++}$, it is informative to analyze other channels with a recoiling $\Delta^{++}$. Equation~\ref{eqn:intenUnstable} can still be used with an appropriate selection of the resonance decay amplitude $Z(\Phi,\Omega)$. The simplest of these channels is $\gamma p\to\pi^-\Delta^{++}$, where preliminary amplitude fits show good agreement with the spin-density matrix elements (SDME) of $\pi^-\Delta^{++}$. Extraction of the SDMEs of the target proton to recoil $\Delta^{++}$ transition provide an independent determination of which amounts of natural or unnatural exchanges are present, as done for the $\Lambda(1520)$ in Ref.~\cite{ref:peter}. This is an important cross-check of this intensity formulation and may provide information as to which baryon spin flip and exchange naturality combinations, for example, are dominant or negligible. 

Further instructive cases include $\gamma p\to\eta'\pi^-\Delta^{++}$ and $\gamma p\to\pi^0\pi^-\Delta^{++}$. Resonances decaying to $\eta'\pi^-$ may include the spin-exotic $\pi_1^-(1600)$~\cite{ref:jpac_pi1,ref:woss}, which would decay in a $P$-wave. Any odd-$L$ contributions to the $\eta'\pi$ channel must have spin-exotic quantum numbers~\cite{ref:compass_etapi}, so observation of a resonant $P$-wave would be clear indication of a spin-exotic state. The dominant contribution to the $\pi^0\pi^-$ channel is the $\rho^-$, which decays in a $P$-wave. Thus, analysis of $\gamma p\to\pi^0\pi^-\Delta^{++}$ will provide a picture of what a resonance decaying to two pseudoscalars in a $P$-wave looks like, and will inform analysis of $\gamma p\to\eta'\pi^-\Delta^{++}$.

\section{Analysis Status and Outlook}
The channel $\gamma p \to \omega\pi^0 p$ is dominated by the axial vector resonance $b_1$, which allows for study of the properties of that meson. The intensity function described in Eqn.~\ref{eqn:intenUnstable} is being tested in multiple channels involving $\Delta^{++}$ recoil at GlueX, in particular, $\gamma p\to\omega\pi^-\Delta^{++}$. Monte Carlo samples are generated using the full unstable recoil model and fit using the full model as well as the stable proton recoil model described in Eqn.~\ref{eqn:intenStable} and an SDME formulation similar to the one used for $\pi^-\Delta^{++}$. Since Eqn.~\ref{eqn:intenUnstable} is an extension of Eqn.~\ref{eqn:intenStable}, it should be returned if the $\Delta^{++}$ decay angles in Eqn.~\ref{eqn:intenUnstable} are integrated over. Likewise, if the resonance angles in Eqn.~\ref{eqn:intenUnstable} are integrated over, the SDME formulation for the baryon vertex is returned, producing a relationship between the amplitude and SDME formulations. Thus, it is expected that fits using the SDME formulation or Eqn.~\ref{eqn:intenStable} will be able to extract at least some correct information from the data, and fits to Monte Carlo samples will indicate if they can. Quantities under investigation include the reflectivity of the reaction, the helicity of the intermediate resonance, and the helicity flip between the target proton and recoil baryon. All of these quantities will be critical in the eventual amplitude analysis.

\acknowledgments
Thanks to F. Afzal, V. Mathieu, M.R. Shepherd, and J.R. Stevens for the many productive discussions involved in this work. This work was funded in part by the US Department of Energy Office of Nuclear Physics under grant DE-FG02-05ER41374.

\end{document}